\documentclass[sn-mathphys-num]{sn-jnl}

\usepackage[utf8]{inputenc}
\usepackage[T1]{fontenc}

\usepackage{multirow}%
\usepackage{mathrsfs}%
\usepackage[title]{appendix}%
\usepackage{xcolor}%
\usepackage{manyfoot}%
\usepackage{booktabs}%
\usepackage{algorithmicx}%
\usepackage{listings}%

\usepackage{amsmath, amsthm, amscd, amsfonts, amssymb, graphicx, color,colortbl, latexsym}
\usepackage{textcomp}
\usepackage{graphicx,subfigure}
\usepackage{array}
\usepackage{caption}

\usepackage{booktabs}

\usepackage{algorithm}
\usepackage{algpseudocode}
\usepackage[utf8]{inputenc}
\usepackage[english]{babel}
\usepackage{extpfeil}
\usepackage{tabularx}
\usepackage{balance}
\usepackage{lineno,hyperref}
\usepackage{lipsum}
\usepackage{soul}
\usepackage{placeins}

\theoremstyle{thmstyleone}%
\theoremstyle{thmstyletwo}%

\theoremstyle{thmstylethree}%

\begin{document}
	
	\title[Rethinking Penetration Testing for AI-Enabled Systems: From Resource Compromise to Behavioral Objective Violation]{Rethinking Penetration Testing for AI-Enabled Systems: From Resource Compromise to Behavioral Objective Violation}
	
	
	\author[1]{\fnm{Mohammad} \sur{Allahbakhsh$^*$}}\email{allahbakhsh@um.ac.ir }
	
	\author[2]{\fnm{Mohammad Hassan} \sur{Bahari}}\email{bahari@sensifai.com}
	\author[2]{\fnm{Moslem} \sur{Attar Raouf}}    \email{moslem@sensifai.com}
	
	\affil*[1]{\orgdiv{Computer Engineering Department}, \orgname{Ferdowsi University of Mashhad}, \orgaddress{\city{Mashhad}, \country{Iran}}}
	
	\affil[2]{\orgname{Sensifai BV}, \orgaddress{ \city{Brussels}, \country{Belgium}}}

\abstract{
Penetration testing traditionally evaluates whether adversaries can exploit weaknesses in software, infrastructure, configurations, or operational controls to achieve security-relevant compromise. This paradigm remains necessary for AI-enabled systems, but it is no longer sufficient. In such systems, adversaries may influence prompts, retrieved content, sensor inputs, training data, memory, tools, or human-AI interaction loops to alter system behavior without directly compromising the underlying infrastructure. This paper reframes penetration testing for AI-enabled systems as objective-driven behavioral evaluation. We define an AI-enabled system as one in which learned models materially influence behavior affecting operational outcomes, and we define AI-enabled penetration as the feasible induction of AI-governed behavior that violates one or more operational objectives under an explicit threat model. This definition preserves conventional penetration testing while extending it to adversarial pathways such as prompt injection, indirect prompt injection, data poisoning, sensor manipulation, retrieval poisoning, tool misuse, and agentic misalignment. We further propose a testing workflow that identifies operational objectives, maps AI-governed behavior, analyzes adversarial influence surfaces, defines behavioral failure criteria, executes scenario-based tests, and reports evidence linking adversarial action to objective violation. A running example involving an AI-enabled security operations center assistant illustrates how penetration may occur through behavioral influence rather than infrastructure compromise. Together, the definitions, workflow, and example provide a technical framework for evaluating adversarial success in deployed AI-enabled systems.
}

\keywords{Penetration Testing,
AI Security,
Trustworthy AI,
Adversarial Machine Learning,
AI-Governed Systems,
Operational Objectives,
Behavioral Security,
Security Evaluation,
Cybersecurity,
System Resilience
}
	
	
	
	\maketitle

\section{Introduction}
\label{sec:introduction}

Consider an AI-enabled security operations center assistant that summarizes alerts, retrieves threat intelligence, recommends severity levels, and invokes predefined response playbooks. During a penetration test, the adversary does not steal credentials, exploit a server, or modify a database. Instead, the adversary plants malicious instructions in content later retrieved by the assistant. If the assistant treats that content as instruction rather than evidence, it may downgrade a high-severity incident and fail to escalate it to a human analyst. Under this threat model, no conventional infrastructure compromise has occurred, but the operational objective of accurate incident triage has failed.

This scenario illustrates a gap in how penetration is usually defined for AI-enabled systems. Penetration testing remains a foundational security practice. Standards and practitioner frameworks such as NIST SP 800-115 and MITRE ATT\&CK provide mature guidance for threat-informed testing, adversary emulation, exploit validation, impact analysis, and mitigation planning \cite{nist800115,mitreattack}. These practices remain necessary because AI-enabled systems still depend on conventional assets, including APIs, credentials, cloud services, model repositories, data stores, containers, orchestration layers, sensors, and software supply chains. If these resources are compromised, the AI-enabled system may be compromised as well.

The difficulty is that resource compromise is no longer the only meaningful path to security failure. In AI-enabled systems, learned models mediate the relationship between system inputs and operational outcomes. Classifiers, large language models, recommender systems, perception models, and agentic workflows may influence decisions, tool calls, prioritization, access decisions, human judgment, or physical actions. As a result, an adversary may affect system behavior through prompts, retrieved content, training data, sensor inputs, context windows, memory, tool outputs, or other ordinary interfaces. Such influence may occur without unauthorized access to the underlying infrastructure or direct control over computational resources.

This shift is visible across several areas of AI security. Adversarial machine learning has shown that learned models can be manipulated through adversarial examples, evasion, poisoning, extraction, inference, and availability attacks \cite{goodfellowAdversarial,biggioWildPatterns,rosenbergAML}. LLM and generative-AI security work has shown that prompt injection, indirect prompt injection, transferable adversarial prompting, insecure tool use, sensitive-information disclosure, and excessive agency can cause harmful behavior through natural-language and workflow interfaces \cite{owaspLLMTop10,duartePromptInjection,greshakeIndirectPromptInjection,zouUniversalTransferable,nistAI6001}. MITRE ATLAS and OWASP AI guidance further organize AI-specific adversary techniques and risks for practitioners \cite{mitreatlas,owaspAIExchange}. These bodies of work make clear that AI security is not merely a question of model accuracy. It is a question of how adversaries can influence AI-mediated behavior in systems that perform operational functions.

However, the meaning of \emph{penetration} in this setting remains underdefined. Existing security-testing frameworks explain how to assess technical systems and adversary behavior. AI risk frameworks such as the NIST AI Risk Management Framework and ISO/IEC 23894 explain how to identify and manage AI risks across organizational contexts \cite{nistAIRMF,iso23894}. AI red-teaming and adversarial ML methods identify important failure modes and attack techniques. Yet these bodies of work do not provide a unified penetration-testing success criterion for cases in which no traditional resource has been compromised, but the system has been induced to act against its operational purpose.

This article addresses that gap by reframing penetration testing around operational objectives and AI-governed behavior. We use \emph{operational objective} to mean a mission-level outcome that a system is expected to preserve under normal and adversarial conditions, such as safe navigation, accurate incident triage, reliable authentication, correct diagnosis, compliant decision support, or trustworthy recommendation. We use \emph{AI-governed behavior} to refer to system behavior materially shaped by learned models, including predictions, generated content, classifications, rankings, plans, tool calls, recommendations, and actions. Under this view, computational resources remain important, but they are not the final object of security evaluation. They enable behavior, and behavior determines whether operational objectives are satisfied.

The central claim of this article is narrow but important: penetration of an AI-enabled system should be understood as the feasible induction of AI-governed behavior that violates one or more operational objectives under an explicit threat model. This definition does not replace conventional penetration testing. It extends its success criterion. Resource compromise remains one possible pathway to penetration, and in many systems it will remain the most important one. But in AI-enabled systems, adversarial influence may also operate through data, prompts, context, sensors, tools, or human-AI interaction loops. A penetration test that ignores these pathways may miss failures that matter operationally.

This framing also avoids a common overstatement. Not every model error is a penetration, and not every harmful AI output is evidence of adversarial success. A behavioral failure becomes penetration only when there is a feasible adversarial path, a defined threat model, an observable AI-governed behavior, and a violation of an operational objective. Because AI behavior may be stochastic, contextual, and distribution-dependent, evidence for penetration may be probabilistic rather than a single deterministic exploit chain. This requires test reports to describe not only the vulnerable interface, but also the adversary capability, induced behavior, success rate or reproducibility conditions, affected objective, operational impact, and recommended mitigation.

The article makes three contributions:

\begin{itemize}
\item It clarifies the conceptual distinction between resource compromise, AI-governed behavior, and operational-objective violation.

\item It defines penetration for AI-enabled systems in a way that preserves the relevance of classical penetration testing while extending its success criterion to behavior-mediated failure.

\item It proposes an objective-driven workflow for AI penetration testing, including how to scope operational objectives, identify adversarial influence surfaces, define behavioral failure criteria, conduct scenario-based testing, and report evidence.

\end{itemize}

The remainder of the paper is organized as follows. Section~\ref{sec:what-pentesting-gets-right} reviews what conventional penetration testing already provides and why its evidentiary discipline remains necessary for AI-enabled systems. Section~\ref{sec:ai-shift} explains the shift from resource compromise to AI-governed behavior. Section~\ref{sec:related-work} positions the paper relative to penetration-testing guidance, adversarial machine learning, LLM and agentic AI security, AI risk management, and cyber-resilience standards. Section~\ref{sec:defining-ai-penetration} introduces the proposed definitions of AI-enabled systems, operational objectives, AI-governed behavior, adversarial influence paths, and AI-enabled penetration. Section~\ref{sec:objective-driven-testing} presents the objective-driven testing workflow. Section~\ref{sec:soc-assistant-example} illustrates the framework using a running example involving manipulation of an AI-enabled SOC assistant. Section~\ref{sec:implications} discusses implications for security practice. Section~\ref{sec:open-challenges} outlines open research challenges, and finally the paper concludes in Section~\ref{sec:conclusion}.

\section{What Penetration Testing Already Gets Right}
\label{sec:what-pentesting-gets-right}

Penetration testing remains one of the most effective ways to evaluate whether security controls withstand adversarial pressure. Its value lies not only in finding vulnerabilities, but in producing evidence that a weakness can be exercised under specified assumptions and can lead to a security-relevant consequence. This distinguishes penetration testing from purely checklist-based audits or vulnerability inventories. A vulnerability report may identify a possible weakness; a penetration test attempts to establish whether that weakness can be used by an adversary to achieve a meaningful objective.

This evidentiary discipline is central to established security-testing practice. NIST SP 800-115 describes technical security testing and assessment as a structured process that includes planning, test execution, analysis of findings, and mitigation guidance \cite{nist800115}. In mature practice, this process requires an explicit scope, defined assumptions, selected techniques, collected evidence, impact assessment, and remediation recommendations. The result is not merely a catalogue of defects, but an adversarial account of what a capable attacker could plausibly do within the boundaries of the test.

Modern penetration testing is also threat-informed. A useful test does not ask whether a system is secure in the abstract; it asks whether an adversary with particular capabilities, access, and constraints can achieve particular objectives. This orientation connects penetration testing to adversary emulation and operational defense. MITRE ATT\&CK provides a widely used vocabulary for describing adversary tactics, techniques, and procedures observed in real environments \cite{mitreattack}. Although ATT\&CK is not itself a penetration-testing methodology, it strengthens security assessment by grounding test activity in realistic adversary behavior rather than arbitrary tool execution.

A further strength of conventional penetration testing is its treatment of exploitability as a matter of evidence. A misconfiguration, exposed service, vulnerable dependency, or scanner finding is not automatically equivalent to compromise. The relevant question is whether the weakness can be exercised under realistic assumptions and whether it enables an adversary to violate a security property. This distinction prevents security assessment from collapsing into vulnerability enumeration. The strongest penetration-test findings connect a weakness to an attack path and connect the attack path to impact.

Conventional penetration testing is also more impact-aware than a narrow caricature suggests. Its findings are typically evaluated in relation to confidentiality, integrity, availability, privilege boundaries, persistence, lateral movement, exposure of sensitive data, operational disruption, or business consequence. This impact orientation also appears in adversarial machine learning. Rosenberg et al. connect attacks against machine-learning systems to conventional cybersecurity objectives, showing how model extraction and inference attacks may affect confidentiality, evasion and poisoning may affect integrity, and resource-exhaustion attacks may affect availability \cite{rosenbergAML}. This continuity matters: AI security is not separate from cybersecurity, but an extension of it into systems whose behavior is partly learned rather than fully specified by code.

The conventional paradigm also remains directly applicable to AI-enabled systems. Such systems depend on ordinary computational resources: APIs, identities, credentials, cloud services, containers, orchestration layers, data stores, model registries, inference endpoints, sensors, logs, software dependencies, and deployment pipelines. If these resources are compromised, the AI-enabled system may be compromised as well. An exposed inference API, overprivileged service account, vulnerable plugin, insecure model repository, poisoned data pipeline, or compromised container can produce familiar security failures. In these cases, conventional penetration testing provides the correct evaluation logic: identify the weakness, exercise an attack path, demonstrate the consequence, and recommend mitigation.

Reproducibility is another important contribution of conventional penetration testing. A strong finding specifies the vulnerable condition, the adversary capability, the steps required to exercise the weakness, the resulting system state, and the conditions under which the attack succeeds. This supports verification by defenders and enables later validation that remediation was effective. AI-enabled systems complicate reproducibility because model outputs may be stochastic, context-dependent, or sensitive to small changes in input. However, this does not reduce the importance of evidentiary rigor. It increases it. AI security assessments still need clear records of test conditions, observed behavior, success frequency, and failure boundaries.

Penetration testing also translates adversarial evidence into defensive action. A finding is useful only if it helps an organization reduce risk. Remediation may involve patching, hardening, segmentation, access-control changes, monitoring improvements, secure-development practices, incident-response changes, or supply-chain controls. The same principle applies to AI-enabled systems. Defenses may involve limiting tool permissions, validating retrieved content, isolating untrusted data, constraining agent actions, monitoring model behavior, hardening pipelines, or introducing human confirmation for high-impact decisions. The defensive value of penetration testing depends on this connection between demonstrated adversarial capability and practical mitigation.

The limitation emerges when security failure is produced without direct compromise of the underlying computational resources. In AI-enabled systems, an adversary may influence behavior through interfaces that are intentionally exposed: a prompt, document, image, sensor reading, retrieved webpage, ticket comment, training example, memory entry, or tool output. The attack may not require unauthorized access to the host, database, model weights, or credentials. The infrastructure may remain intact, while the AI-mediated behavior violates the purpose for which the system was deployed.

This creates a boundary condition for conventional success criteria. Classical penetration testing is strongest when security failure is evidenced by unauthorized access, privilege escalation, data exposure, persistence, disruption, or control over resources. AI-enabled systems introduce another form of adversarial evidence: induced behavior that violates an operational objective. A model may misclassify an object, an agent may misuse an authorized tool, a recommender may steer a user toward an unsafe decision, or a decision-support system may suppress information needed for correct judgment. In each case, the security-relevant failure lies not only in the state of computational resources, but in the behavior those resources produce.

The required extension is therefore conceptual rather than oppositional. Conventional penetration testing supplies the discipline: explicit threat models, adversarial evidence, exploitability analysis, impact assessment, reproducibility, and remediation. AI-enabled systems require that this discipline be applied not only to resources, but also to AI-governed behavior. When learned models mediate the relationship between system inputs and operational outcomes, penetration cannot be defined solely as entry into, control over, or misuse of computational assets. It must also account for adversarial influence over behavior that causes the system to violate its operational objectives.

\section{The AI Shift: From Resources to Behavior}
\label{sec:ai-shift}

The defining change introduced by AI-enabled systems is not simply the addition of a new software component. It is the insertion of learned behavior into the causal path between system resources and operational outcomes. In conventional systems, resources such as files, processes, credentials, memory, network services, and databases are often closely coupled to system behavior. Unauthorized control over these resources frequently enables direct manipulation of execution, data, or availability. In AI-enabled systems, this coupling is weakened. Resources still matter, but their security significance increasingly depends on how they shape AI-governed behavior.

This shift can be understood as a three-layer relationship. The first layer consists of computational resources: infrastructure, code, model artifacts, prompts, APIs, data stores, sensors, credentials, tool interfaces, and deployment pipelines. The second layer consists of AI-governed behavior: predictions, classifications, generated responses, rankings, plans, recommendations, tool calls, and actions materially shaped by learned models. The third layer consists of operational objectives: the mission-level outcomes the system is expected to preserve, such as accurate incident triage, reliable authentication, safe navigation, correct diagnosis, compliant decision support, or trustworthy content moderation. Security failure may occur at any layer, but in AI-enabled systems the second layer becomes a first-class object of adversarial influence.

This is a structural change. In traditional software, behavior is primarily specified by program logic. Although systems may be complex, distributed, and difficult to analyze, their intended behavior is usually encoded in explicit instructions written by developers and constrained by access-control mechanisms, input validation, type systems, protocol definitions, and configuration rules. Penetration testing can therefore often reason from resource compromise to behavioral consequence. If an attacker obtains administrative credentials, modifies a database, controls a process, or exploits a remote-code-execution vulnerability, the path from resource compromise to security impact is relatively direct.

AI-enabled systems introduce an additional mechanism for producing behavior. A learned model does not merely execute explicit rules; it maps inputs, context, and learned representations to outputs. This mapping may be probabilistic, distribution-dependent, context-sensitive, and difficult to specify exhaustively. A classifier may behave differently under small perturbations of an input. A generative model may produce different outputs under different prompts, sampling settings, retrieved context, or conversation history. An agent may select different tools or plans depending on intermediate observations. The resulting behavior is not independent of software logic, but it is not reducible to software logic alone.

This has direct consequences for security assessment. An adversary may influence an AI-enabled system by acting on the inputs, context, data, or environment from which the model derives behavior. In an image-recognition system, adversarial perturbations may alter classification without modifying the model or compromising the host, a phenomenon established in foundational work on adversarial examples and later systematized in adversarial machine learning research~\cite{goodfellowAdversarial,biggioWildPatterns}. In a retrieval-augmented generation system, poisoned or maliciously crafted retrieved content may steer the generated answer without changing the underlying model weights. In an agentic system, an indirect prompt injection may cause the agent to misuse an authorized tool even though the tool, account, and infrastructure remain uncompromised. In a cyber-physical system, sensor manipulation may affect perception and planning without breaching the control server.

These pathways differ from classical exploitation because the adversary's leverage is behavioral rather than primarily structural. The attacker does not necessarily need to cross an authorization boundary, escalate privileges, or obtain persistent access. The attack may proceed through an interface the system intentionally exposes: a prompt, document, ticket, webpage, image, audio stream, sensor reading, memory item, API response, or training example. The resulting failure is not always visible as a compromised asset. It may appear as a wrong recommendation, unsafe classification, unauthorized tool call, suppressed warning, distorted ranking, misleading summary, or policy-violating action.

Large language models and agentic AI systems make this problem especially visible. Their inputs often combine system instructions, developer instructions, user messages, retrieved documents, tool outputs, memory, and external content in a shared natural-language context. Prompt injection exploits this mixture by causing untrusted content to influence model behavior \cite{owaspLLMTop10,duartePromptInjection}. Indirect prompt injection extends the attack surface further by placing adversarial instructions in content that the model later retrieves or observes, such as webpages, documents, emails, tickets, or database records \cite{greshakeIndirectPromptInjection}. The attacker may never interact directly with the model at the time of execution; the environment becomes the delivery mechanism for adversarial influence. Related work on transferable adversarial attacks against aligned language models further shows that adversarial prompting can systematically alter model behavior across systems \cite{zouUniversalTransferable}.

The confused-deputy pattern captures one important form of this risk. In the classical formulation, a privileged program is tricked into misusing its authority on behalf of an adversary \cite{hardyConfusedDeputy}. In agentic AI, the model or agent may become the deputy. It may have legitimate access to tools, files, APIs, or workflow actions, while untrusted content shapes its decision about how to use those capabilities. The adversary's goal is not necessarily to steal the agent's credentials or compromise the tool. It is to induce the agent to exercise legitimate authority in a way that violates the operator's intent. This turns behavior itself into the target of penetration.

The same pattern appears beyond LLMs. In biometric authentication, an adversarially crafted presentation may cause a system to accept an unauthorized user without compromising the authentication server. In autonomous inspection, a manipulated visual input may cause a defect-detection model to miss a critical fault while all infrastructure controls remain intact. In recommender systems, coordinated manipulation of inputs may shift rankings or recommendations in ways that distort user decisions or operational priorities. In medical or security decision support, carefully crafted context may produce advice that is plausible, harmful, and operationally consequential. These cases differ technically, but they share the same structure: adversarial influence over AI-governed behavior produces objective-level failure.

This structure also changes how success should be evaluated. In conventional penetration testing, success is often demonstrated by a reproducible exploit chain leading to a discrete state change: a shell is obtained, a privilege is escalated, a file is accessed, a service is disrupted, or data is exfiltrated. AI-enabled systems may not always produce such binary evidence. A behavioral attack may succeed with some probability across repeated trials. It may depend on prompt phrasing, retrieval order, environmental state, sensor conditions, model version, sampling parameters, memory contents, or human interaction. Evidence of penetration may therefore require repeated scenario execution, behavioral observation, success-rate estimation, and explicit documentation of test conditions.

The probabilistic nature of AI behavior does not make testing less rigorous. It changes the form of rigor required. A single surprising output from a model is weak evidence unless it is connected to an adversarial path and an operational consequence. Conversely, repeated induction of harmful behavior under defined conditions may constitute strong evidence even if the behavior is not deterministic. The relevant evidentiary question is whether an adversary with specified capabilities can reliably enough induce behavior that violates an operational objective. For high-impact systems, even a moderate probability of induced objective failure may be security-relevant if the operational consequence is severe.

This shift also affects the meaning of vulnerability. In a conventional setting, vulnerability is often associated with a defect in code, configuration, protocol design, access control, or deployment. In an AI-enabled setting, a vulnerability may instead be a behavioral susceptibility: a condition under which untrusted input, context, data, or environmental manipulation can steer AI-governed behavior outside the acceptable envelope for the system's objective. The weakness may lie in the model, but it may also lie in prompt construction, retrieval design, tool authorization, memory handling, workflow orchestration, monitoring, or the allocation of decision authority between the AI system and human operators.

The boundary between resources and behavior should not be overstated. Resource compromise remains a powerful route to behavioral compromise. If an attacker modifies model weights, tampers with training data, steals credentials, poisons a retrieval index, or controls an agent's tool interface, the resulting behavioral consequences may be severe. The point is not that resources have become irrelevant. The point is that resources are no longer sufficient as the sole object of penetration analysis. AI-enabled systems require testers to examine both the protection of resources and the integrity of the behavior those resources produce.

This motivates an objective-centered view of security assessment. The decisive question is not only whether an attacker can enter, control, or misuse a computational resource. It is whether the attacker can cause the AI-enabled system to act contrary to the operational purpose for which it was deployed. When learned models mediate the relationship between resources and outcomes, penetration testing must account for the full path from adversarial influence to AI-governed behavior to operational-objective violation.

\section{Related Work and Standards}
\label{sec:related-work}

Prior work on penetration testing, adversarial machine learning, LLM security, AI red teaming, and AI risk management provides essential foundations for evaluating AI-enabled systems. These bodies of work address important parts of the problem, but they do not fully define what it means to penetrate a system whose critical behavior is mediated by learned models. The gap is not a lack of security guidance for AI. The gap is the absence of a clear success criterion for penetration testing when adversarial influence causes AI-governed behavior to violate operational objectives without necessarily compromising computational resources.

\subsection{Security Testing and Adversary Modeling}

Conventional security-testing guidance provides the procedural foundation for penetration testing. NIST SP 800-115 defines technical security testing and assessment as a structured activity involving planning, execution, analysis, and mitigation guidance \cite{nist800115}. This framing is important because penetration testing is not merely vulnerability discovery. It is a process for generating adversarial evidence under defined assumptions and translating that evidence into risk-reduction actions. MITRE ATT\&CK complements this process by providing a widely used vocabulary of adversary tactics, techniques, and procedures observed in real environments \cite{mitreattack}. Together, these frameworks support threat-informed testing, adversary emulation, attack-path analysis, and defensive prioritization.

These frameworks remain directly relevant to AI-enabled systems. AI applications still rely on ordinary infrastructure, including APIs, identities, cloud services, model repositories, containers, data stores, sensors, and software supply chains. Attacks against those resources remain within the scope of conventional penetration testing. However, these frameworks do not provide a specific definition of penetration for cases in which adversarial success occurs through AI-mediated behavior rather than direct compromise of resources. They describe how adversaries operate and how tests should be conducted, but they do not fully specify how to treat induced model or agent behavior as the object of penetration.

\subsection{Adversarial Machine Learning}

Adversarial machine learning provides the first major body of work on intentional manipulation of learned models. Foundational work on adversarial examples showed that small, carefully crafted perturbations can change model predictions, making model behavior itself an object of adversarial control~\cite{goodfellowAdversarial}. Broader adversarial machine learning research has since organized attacks and defenses around threat models, attacker knowledge, attack surfaces, and evaluation methodology~\cite{biggioWildPatterns}. This literature studies evasion attacks, poisoning attacks, model extraction, membership inference, model inversion, privacy leakage, and resource-exhaustion attacks. Rosenberg et al. connect these attacks to conventional cybersecurity objectives by showing how attacks on ML systems can affect confidentiality, integrity, and availability \cite{rosenbergAML}. This connection is important because it prevents AI security from being reduced to accuracy degradation alone. ML security failures can expose sensitive information, corrupt decisions, or degrade availability in ways that align with established cybersecurity impact categories.

OWASP's Machine Learning Security Top 10 and MITRE ATLAS further translate AI-specific threats into practitioner-oriented taxonomies \cite{owaspMLTop10,mitreatlas}. OWASP organizes recurring security risks in ML systems, while MITRE ATLAS catalogues adversarial tactics and techniques against AI-enabled systems. These resources are valuable because they extend security assessment beyond traditional software vulnerabilities and make AI-specific attack surfaces visible to practitioners.

The limitation is that adversarial ML is often model-centric or pipeline-centric. A model may be evaluated as vulnerable if an adversarial input changes its classification, if poisoned training data changes its learned behavior, or if an adversary can infer information about its training data. These are real security concerns, but they do not always determine whether the larger AI-enabled system has been penetrated. A model error may be operationally irrelevant in one context and mission-critical in another. Conversely, an adversary may cause operational failure by manipulating a workflow, retrieval context, tool invocation, or human-AI interaction without fitting neatly into a single model-level attack category. This motivates a system-level definition that connects adversarial influence to operational-objective violation.

\subsection{LLM and Agentic AI Security}

LLM and generative-AI systems make the behavioral nature of AI security especially visible. OWASP's Top 10 for LLM and Generative AI Applications identifies risks such as prompt injection, insecure output handling, sensitive-information disclosure, model denial of service, supply-chain vulnerabilities, and excessive agency \cite{owaspLLMTop10}. NIST AI 600-1, the Generative AI Profile of the NIST AI Risk Management Framework, similarly highlights risks associated with generative AI systems across design, development, deployment, and use \cite{nistAI6001}. The OWASP AI Exchange provides broader guidance for AI security and privacy, including threats, controls, and risk-analysis practices for AI and data-centric systems \cite{owaspAIExchange}.

A central issue in this literature is the boundary between instructions and data. Prompt injection and indirect prompt injection exploit the fact that LLM-based systems may process system instructions, developer instructions, user input, retrieved documents, tool outputs, and external content within a shared context. Duarte et al. characterize prompt injection as a systematic class of attacks that can influence model behavior through natural-language inputs and tool-connected workflows \cite{duartePromptInjection}. Greshake et al. show that indirect prompt injection can compromise LLM-integrated applications by placing adversarial instructions in external data that the application later retrieves or processes \cite{greshakeIndirectPromptInjection}. Work on transferable adversarial attacks against aligned language models further demonstrates that adversarial prompting can be automated and transferred across systems \cite{zouUniversalTransferable}. This does not mean that instruction--data separation is impossible. Work on Instruction Hierarchy and StruQ explicitly attempts to restore authority ordering or structural separation between trusted instructions and untrusted data \cite{instructionHierarchy,struq}. The more defensible conclusion is that current LLM-integrated systems make this separation fragile, context-dependent, and difficult to validate across multi-step workflows.

Agentic AI extends the problem because the model is no longer limited to generating text. Agents may retrieve information, call tools, update memory, interact with APIs, modify records, initiate workflows, or make recommendations that shape human decisions. Kim et al. analyze agentic AI security through design dimensions such as tool integration, access sensitivity, autonomy, and data flow, showing how these dimensions affect the attack surface \cite{kimAgenticAI}. Deng et al. similarly identify security challenges across the agent lifecycle, including planning risk, tool misuse, and misalignment between intended and actual behavior \cite{dengAgentLifecycle}. These studies support the view that the security object is not only the model, but the behavior of the model within a larger operational loop.

The confused-deputy problem provides a classical bridge to this setting. Hardy's original formulation describes a privileged program that is tricked into misusing its authority on behalf of an adversary \cite{hardyConfusedDeputy}. In agentic AI, the agent may become the deputy. It may have legitimate access to tools, files, APIs, credentials, or workflow actions, while untrusted content influences how it uses those capabilities. The adversary may not need to steal credentials or compromise the tool. Instead, the adversary induces the agent to exercise legitimate authority in a way that violates the system operator's intent. This pattern shows why AI penetration cannot be defined solely by unauthorized access to resources.

\subsection{AI Risk Management and Cyber Resilience}

AI risk-management standards provide a broader governance context for this discussion. The NIST AI Risk Management Framework defines a structured approach for mapping, measuring, managing, and governing AI risks \cite{nistAIRMF}. ISO/IEC 23894 provides guidance for integrating AI risk management into organizational processes \cite{iso23894}. NIST SP 800-160 Volume 2 Revision 1 frames cyber resilience from a systems-security-engineering perspective, emphasizing the ability of systems to anticipate, withstand, recover from, and adapt to adverse conditions \cite{nist800160v2r1}. These frameworks are important because they already move beyond isolated vulnerabilities toward system context, risk, resilience, and organizational objectives.

However, risk-management frameworks and penetration-testing definitions serve different purposes. NIST AI RMF and ISO/IEC 23894 help organizations identify and manage AI risks, but they do not define penetration success in adversarial testing. NIST SP 800-160 provides a resilience-oriented systems-engineering foundation, but it is not an AI penetration-testing methodology. MITRE ATLAS and OWASP guidance identify attack techniques and control areas, but they do not fully specify when induced AI behavior should count as successful penetration of an AI-enabled system. These frameworks are complementary to the present work, but they leave open the conceptual boundary between AI risk, AI red teaming, and penetration success.

\begin{table*}[t]
\centering
\caption{Positioning of the proposed definition relative to selected standards and related work.}
\label{tab:related-work-positioning}
\renewcommand{\arraystretch}{1.15}
\begin{tabular}{p{0.20\textwidth} p{0.22\textwidth} p{0.27\textwidth} p{0.23\textwidth}}
\hline
\textbf{Framework or literature} & \textbf{Primary focus} & \textbf{What it covers well} & \textbf{Remaining gap for AI penetration testing} \\
\hline
NIST SP 800-115 \cite{nist800115} & Technical security testing and assessment & Planning, executing, analyzing, and reporting security tests & Does not define penetration for AI-mediated behavioral failure \\
\hline
MITRE ATT\&CK \cite{mitreattack} & Adversary tactics and techniques & Threat-informed adversary emulation and attack behavior & Not centered on AI-governed behavior or operational-objective violation \\
\hline
Adversarial ML \cite{goodfellowAdversarial,biggioWildPatterns,rosenbergAML} & Attacks and defenses for ML systems & Adversarial examples, evasion, poisoning, extraction, inference, and CIA-oriented impact & Often model- or pipeline-centric rather than operational-objective-centric\\
\hline
MITRE ATLAS \cite{mitreatlas} & AI-specific adversary tactics and techniques & Taxonomy of attacks against AI-enabled systems & Catalogues techniques but does not define penetration success at mission level \\
\hline
LLM and agentic AI security \cite{owaspLLMTop10,greshakeIndirectPromptInjection,zouUniversalTransferable,kimAgenticAI,dengAgentLifecycle} & LLM and agentic-AI application risks & Prompt injection, indirect prompt injection, excessive agency, adversarial prompting, tool misuse, and agent lifecycle risks & Identifies behavioral attack paths but does not define penetration success as operational-objective violation \\
\hline
NIST AI RMF and ISO/IEC 23894 \cite{nistAIRMF,iso23894} & AI risk management & Governance, risk mapping, measurement, management, and organizational context & Address AI risk broadly but not penetration-testing success criteria \\
\hline
NIST SP 800-160 Vol. 2 Rev. 1 \cite{nist800160v2r1} & Cyber resilience and systems security engineering & Mission-aware resilience, survivability, and adaptation under adverse conditions & Supports objective-level thinking but is not AI-specific penetration guidance \\
\hline
\end{tabular}
\end{table*}
\FloatBarrier

\subsection{Gap and Positioning}

The literature establishes several important points. Conventional penetration testing provides adversarial evidence, exploitability analysis, and mitigation guidance. Threat-informed frameworks describe realistic adversary behavior. Adversarial ML explains how learned models and ML pipelines can be attacked. LLM and agentic AI security research shows how prompts, tools, memory, retrieval, and autonomy create new pathways for adversarial influence. AI risk and resilience standards connect AI systems to governance, organizational risk, and mission context.

What remains underdefined is the meaning of penetration itself when adversarial success is achieved through AI-governed behavior. Existing work can identify a prompt injection, a poisoning attack, a model-extraction risk, an unsafe output, or an agentic tool-use failure. It can also describe the organizational risk associated with such failures. Yet these resources do not provide a unified penetration-testing criterion for cases where the adversary does not directly compromise infrastructure but induces behavior that violates the system's operational purpose.

The contribution of this article is to fill that conceptual gap. It defines penetration of an AI-enabled system as the feasible induction of AI-governed behavior that violates one or more operational objectives under an explicit threat model. This definition preserves the relevance of conventional penetration testing and adversarial ML while extending their success criteria to behavior-mediated failure. Resource compromise remains a central path to penetration, but it is not the only one. In AI-enabled systems, the decisive security evidence may be the demonstrated ability to steer behavior away from the objective the system is intended to preserve.

\section{Defining AI-Enabled Penetration}
\label{sec:defining-ai-penetration}

The preceding sections establish that AI-enabled systems require a broader object of penetration testing than computational resources alone. Conventional penetration testing remains valid when adversarial success depends on unauthorized access, privilege escalation, data exposure, disruption, persistence, or control over infrastructure. AI-enabled systems add another possibility: adversarial success may be achieved by influencing AI-governed behavior so that the system violates the operational objective it was deployed to preserve. A precise definition of AI-enabled penetration must therefore connect three elements: the adversary's feasible influence path, the AI-governed behavior induced by that path, and the operational objective violated as a result.

\subsection{AI-Enabled Systems}

An AI-enabled system is not merely a software system that contains a model. Many deployed models support offline analytics or auxiliary functions without materially shaping operational outcomes. The relevant class of systems is narrower: systems in which learned models influence decisions, recommendations, actions, classifications, rankings, or interactions that affect the system's operational purpose.

\noindent\textbf{Definition 1: AI-enabled system.}
An \emph{AI-enabled system} is a system in which one or more learned models materially influence system behavior that affects operational outcomes.

This definition includes systems based on classifiers, perception models, recommender systems, generative models, biometric models, anomaly detectors, decision-support models, and agentic AI workflows. It also includes hybrid systems in which AI components interact with conventional software, databases, APIs, sensors, tools, human operators, or cyber-physical processes. The definition excludes cases where AI is incidental to the security-relevant behavior of the system.

\subsection{Operational Objectives}

The operational objective is the mission-level outcome that gives security significance to system behavior. A facial-recognition system may have the objective of admitting only authorized users. A security operations assistant may have the objective of correctly triaging and escalating high-severity incidents. A medical decision-support system may have the objective of preserving clinically appropriate recommendations. An industrial inspection system may have the objective of detecting safety-critical defects. These objectives are not identical to the resources used to implement the system. They describe what the system is expected to accomplish.

\noindent\textbf{Definition 2: Operational objective.}
An \emph{operational objective} is a mission-level outcome or constraint that an AI-enabled system is expected to preserve under normal and adversarial conditions.

Operational objectives may be binary, threshold-based, probabilistic, or context-dependent. In some systems, the objective may be expressed as a hard constraint, such as never executing a destructive tool action without human authorization. In others, it may be expressed as a performance or safety threshold, such as maintaining incident-prioritization accuracy above an acceptable level under adversarial input conditions. In high-impact settings, operational objectives should be defined before testing begins, because the same AI behavior may have different security significance in different operational contexts.

\subsection{AI-Governed Behavior}

AI-governed behavior is the mediating layer between system resources and operational objectives. It is the behavior produced or materially shaped by learned models within the larger system. This includes not only model outputs, but also downstream decisions and actions influenced by those outputs.

\noindent\textbf{Definition 3: AI-governed behavior.}
\emph{AI-governed behavior} is system behavior materially shaped by learned models, including predictions, classifications, generated content, rankings, recommendations, plans, tool calls, access decisions, alerts, summaries, or actions.

This definition is intentionally system-level. A model prediction is part of AI-governed behavior, but it is not always the entire behavior of interest. In a retrieval-augmented assistant, the relevant behavior may include document retrieval, answer generation, citation selection, tool invocation, and escalation recommendation. In an autonomous inspection workflow, it may include perception, anomaly scoring, routing decisions, and human notification. In an agentic system, it may include planning, memory use, API selection, and execution of authorized actions. Penetration testing must therefore observe the behavior of the AI component within the system, not only the isolated model output.

\subsection{Influence Paths}

An adversary can influence AI-governed behavior through different paths. Some paths involve conventional resource compromise: stealing credentials, modifying model weights, tampering with a database, compromising a deployment pipeline, or gaining control over a tool interface. Other paths operate through interfaces that are intentionally exposed: prompts, documents, webpages, sensor inputs, logs, emails, tickets, retrieved records, training examples, memory entries, or API responses. Both types of path may be security-relevant. The distinction is that AI-enabled systems can produce serious operational failure even when the adversary does not directly compromise the resources that execute the system.

\noindent\textbf{Definition 4: Adversarial influence path.}
An \emph{adversarial influence path} is a feasible sequence of adversary actions, available under a specified threat model, that affects AI-governed behavior through resources, inputs, context, data, environment, tools, memory, or human-AI interaction.

The influence path must be feasible under the assumed threat model. A claim of penetration is weak if it assumes unrealistic adversary access, ignores deployment constraints, or depends on arbitrary manipulation unavailable to the attacker. For example, changing model weights may be a valid influence path if the adversary has compromised the model repository, but not if the threat model grants only external user access. Conversely, placing adversarial instructions in a webpage may be a valid influence path for an agent that routinely retrieves and follows content from the web. The legitimacy of the penetration claim depends on the match between adversary capability and system exposure.

\subsection{Penetration in AI-Enabled Systems}

The central definition follows from the relationship among influence path, behavior, and objective. Penetration occurs when an adversary can feasibly induce AI-governed behavior that violates an operational objective. The definition is intentionally broader than resource compromise, but not broader than adversarial success. It does not classify every model error, hallucination, misclassification, or undesirable output as penetration. The behavior must be induced through an adversarial path and must violate an operational objective under the stated threat model.

\noindent\textbf{Definition 5: AI-enabled penetration.}
\emph{AI-enabled penetration} is the feasible induction of AI-governed behavior that violates one or more operational objectives under an explicit threat model.

This definition generalizes conventional penetration testing rather than replacing it. In a conventional system, penetration may be demonstrated by compromising resources in a way that violates confidentiality, integrity, or availability. In an AI-enabled system, resource compromise remains one possible influence path. However, penetration may also be demonstrated through behavioral influence that does not require direct control over infrastructure. Prompt injection, indirect prompt injection, adversarial perturbation, data poisoning, sensor manipulation, retrieval poisoning, memory manipulation, and tool-output manipulation may all become penetration paths when they induce behavior that violates the operational objective of the system.

The definition also separates \emph{model failure} from \emph{system penetration}. A classifier error alone is not necessarily penetration. A harmful LLM response alone is not necessarily penetration. A failed recommendation alone is not necessarily penetration. These failures become penetration only when they are induced by an adversary with feasible capabilities and when the resulting behavior violates an operational objective. This distinction is essential because AI systems may fail for many non-adversarial reasons, including distribution shift, insufficient training data, ambiguity, noise, poor calibration, or ordinary model uncertainty. Penetration testing is concerned with adversarially induced failure, not failure in general.

\subsection{Successful Penetration}

Successful penetration requires evidence. In conventional testing, evidence may take the form of a reproducible exploit chain, an obtained shell, access to restricted data, privilege escalation, or service disruption. In AI-enabled systems, evidence may take a different form because the induced behavior may be probabilistic, context-dependent, or sensitive to the system state. A single example may be sufficient in some high-impact cases, but stronger evidence often requires repeated trials, controlled conditions, and documentation of success frequency.

\noindent\textbf{Definition 6: Successful AI-enabled penetration.}
\emph{Successful AI-enabled penetration} occurs when a tester demonstrates, under a specified threat model, a feasible adversarial influence path that induces AI-governed behavior resulting in violation of an operational objective.

This definition emphasizes demonstration rather than speculation. A tester must identify the adversary capability, the interface or resource used for influence, the induced behavior, the operational objective violated, and the conditions under which the violation occurs. When the behavior is stochastic, the evidence should include the number of trials, success rate, relevant model or system configuration, prompt or input conditions, retrieval state, tool permissions, and environmental assumptions. The result need not always be deterministic, but it must be sufficiently evidenced to support a security claim.

\subsection{Objective Violation}

Operational-objective violation provides the criterion that distinguishes security-relevant behavioral influence from ordinary model imperfection. The violation may be direct or indirect. A direct violation occurs when the AI-enabled system itself performs an unauthorized, unsafe, or policy-violating action. An indirect violation occurs when the system materially shapes a downstream decision, workflow, or human judgment in a way that defeats the operational purpose of the system.

\noindent\textbf{Definition 7: Operational-objective violation.}
An \emph{operational-objective violation} occurs when AI-governed behavior causes, enables, or materially contributes to a failure of a defined operational objective under the conditions of the test.

This definition allows different evidence standards for different systems. In an access-control system, violation may be unauthorized admission. In a SOC assistant, it may be failure to escalate a high-severity incident. In a medical decision-support system, it may be recommendation of an unsafe action under adversarially manipulated context. In a content-moderation system, it may be systematic failure to enforce a defined policy. In an autonomous system, it may be unsafe planning or control behavior. The operational objective determines what counts as failure; the threat model determines whether the failure counts as penetration.

\subsection{Relationship to Conventional Penetration}

Conventional penetration testing can be viewed as a special case of this broader definition. In many systems, compromising resources is the most direct way to violate an operational objective. If an attacker gains administrative access, modifies a database, disables a service, or exfiltrates sensitive data, the path from resource compromise to objective violation may be straightforward. Such cases remain fully covered by the proposed definition because resource compromise is an adversarial influence path that can induce security-relevant behavior or state changes.

The broader definition becomes necessary when the adversary's influence operates through behavior rather than direct resource control. An attacker who causes an AI assistant to suppress an alert, an agent to call an unauthorized tool, a perception system to miss a safety-critical object, or a recommender system to distort operational priorities may achieve a meaningful security outcome without conventional infrastructure compromise. In these cases, the system has not necessarily been penetrated in the classical sense of unauthorized entry, but it has been penetrated in the operational sense that adversarial influence has defeated the behavior required for the system's mission.

This distinction leads to a compact characterization:

\begin{quote}
An AI-enabled system is penetrated when an adversary can feasibly make the system act against its operational purpose through AI-governed behavior.
\end{quote}

The characterization is not a replacement for the formal definition, because the formal definition requires an explicit threat model, an influence path, induced behavior, and objective violation. It captures the central shift: penetration testing for AI-enabled systems must evaluate not only whether an attacker can get into the system, but whether the attacker can make the system behave in a way that defeats the purpose for which it was deployed.

\section{Objective-Driven AI Penetration Testing}
\label{sec:objective-driven-testing}

The definition of AI-enabled penetration changes the structure of the test. A conventional test often begins with assets, exposure, vulnerabilities, and exploit paths. An objective-driven AI penetration test begins with the operational purpose of the system and works backward to the AI-governed behaviors that preserve or violate that purpose. The central question is not only whether an attacker can compromise a resource, but whether an attacker can induce behavior that defeats the objective the system is expected to maintain.

This does not eliminate asset-centric testing. Infrastructure, identities, APIs, data stores, model artifacts, and deployment pipelines remain part of the test scope. The difference is that they are evaluated together with the behavioral layer that connects them to operational outcomes. An objective-driven test therefore treats resources, AI-governed behavior, and operational objectives as a single adversarial chain.

\subsection{Step 1: Define Operational Objectives}

The first step is to define the operational objectives that the AI-enabled system is expected to preserve. These objectives must be specific enough to support testing. General statements such as `the system should be safe,'' `the model should be reliable,'' or ``the assistant should be trustworthy'' are insufficient. A testable operational objective specifies the behavior or outcome that must hold under adversarial conditions.

For example, in an AI-enabled security operations center assistant, an objective may be: high-severity incidents must not be downgraded or closed without human analyst confirmation. In a biometric access-control system, the objective may be: unauthorized users must not be admitted through adversarial presentation or input manipulation. In an autonomous inspection system, the objective may be: safety-critical defects must be escalated when they exceed a defined confidence or severity threshold. In an LLM-based compliance assistant, the objective may be: regulated decisions must not be recommended without citing approved policy sources.

Operational objectives may be expressed as hard constraints, thresholds, policies, or probabilistic performance requirements. The essential requirement is that the objective defines what counts as failure in the context of the system's mission. Without this definition, the test cannot distinguish between an ordinary model error, a harmless anomaly, and a security-relevant penetration outcome.

\subsection{Step 2: Map AI-Governed Behavior}

The second step is to identify the AI-governed behaviors that influence each operational objective. This includes the model outputs and the downstream behaviors shaped by those outputs. In many systems, the security-relevant behavior is not the raw model prediction, but the decision, recommendation, action, or workflow change produced after the model output is consumed by other components.

For an LLM agent, relevant behaviors may include answer generation, retrieval selection, tool invocation, memory updates, API calls, escalation decisions, and human-facing recommendations. For a perception system, relevant behaviors may include classification, detection confidence, object tracking, anomaly scoring, and downstream control actions. For a recommender system, relevant behaviors may include ranking, filtering, prioritization, and presentation to users. For a decision-support system, relevant behaviors may include evidence selection, risk scoring, explanation generation, and recommendation of actions.

Mapping AI-governed behavior clarifies where adversarial influence can alter the path from inputs to outcomes. It also prevents the test from focusing too narrowly on isolated model outputs. A model may produce a flawed output that is neutralized by downstream controls, or it may produce a subtle output that becomes severe only after aggregation, ranking, tool use, or human interpretation. Objective-driven testing therefore observes the system-level behavior that affects the operational objective.

\subsection{Step 3: Identify Adversarial Influence Surfaces}

The third step is to identify the interfaces through which an adversary can influence AI-governed behavior under the assumed threat model. These influence surfaces may include conventional resources and AI-specific behavioral inputs. Conventional surfaces include exposed services, APIs, credentials, containers, cloud resources, software dependencies, databases, model registries, deployment pipelines, and monitoring systems. AI-specific surfaces include prompts, retrieved documents, emails, tickets, webpages, images, audio streams, sensor readings, training examples, memory entries, tool outputs, logs, and user feedback.

The influence surface must be evaluated relative to adversary capability. A public user may be able to submit prompts, upload files, create content later retrieved by the system, or interact with an exposed API. An insider may be able to modify a knowledge base, annotate training data, influence feedback loops, or access internal workflow tools. A supply-chain adversary may be able to affect model dependencies, datasets, plugins, or external services. A physical adversary may be able to manipulate sensors, signs, objects, lighting, sound, or environmental conditions.

This step produces the adversarial influence map: a structured account of how an attacker can reach the behavioral layer of the system. The map should include both direct and indirect pathways. A direct pathway occurs when the attacker interacts with the AI component itself. An indirect pathway occurs when the attacker modifies or places content that the AI component later observes, retrieves, summarizes, classifies, or acts upon.

\subsection{Step 4: Define Behavioral Failure Criteria}

The fourth step is to translate each operational objective into behavioral failure criteria. These criteria specify what induced behavior counts as objective violation. They must be concrete enough to support evidence collection and severity assessment.

Behavioral failure criteria may include unauthorized action, unsafe recommendation, incorrect escalation, policy bypass, malicious or unintended tool invocation, suppression of critical information, systematic misclassification, harmful ranking shift, misleading summary, privacy-violating disclosure, or manipulation of human decision-making. In agentic systems, failure criteria may also include unauthorized memory modification, execution of a prohibited tool, failure to request required confirmation, or use of untrusted content as authoritative instruction.

The criteria should distinguish severity levels. A low-severity behavioral deviation may produce an incorrect but harmless output. A medium-severity deviation may mislead a user while remaining reversible. A high-severity deviation may cause unauthorized action, delayed response, unsafe control behavior, regulatory violation, or material operational harm. Severity depends on the objective, the system's authority, the affected workflow, and the availability of compensating controls.

Behavioral failure criteria also prevent overclaiming. A single hallucinated statement, incorrect classification, or unexpected output does not automatically establish penetration. The behavior must be induced by an adversary through a feasible influence path and must violate a defined operational objective. This evidentiary threshold separates objective-driven penetration testing from general model evaluation.

\subsection{Step 5: Execute Scenario-Based Tests}

The fifth step is to execute adversarial scenarios that connect influence surfaces to behavioral failure criteria. Scenario-based testing is necessary because AI-enabled failures often depend on context, workflow state, model configuration, tool permissions, retrieval contents, sampling parameters, memory, or human interaction. A test case should therefore specify the adversary capability, initial system state, input or manipulation, expected behavioral effect, observed behavior, and objective violation.

For prompt-based systems, scenarios may test direct prompt injection, indirect prompt injection, conflicting instructions, untrusted retrieved content, malicious tool outputs, or attempts to override policy constraints. For perception systems, scenarios may test adversarial perturbations, environmental manipulation, sensor spoofing, occlusion, lighting changes, or distribution shifts. For training or adaptation pipelines, scenarios may test poisoning, biased feedback, malicious labeling, data contamination, or model-update manipulation. For agentic systems, scenarios may test tool misuse, planning manipulation, excessive agency, memory poisoning, authorization bypass, or confused-deputy behavior.

Because AI behavior may be stochastic, scenario execution may require repeated trials. The number of trials depends on the system's impact, the variability of behavior, and the severity of the objective violation. Evidence may include success frequency, confidence intervals, representative transcripts, model outputs, tool logs, retrieval traces, system state snapshots, and human-review outcomes. Deterministic reproduction is valuable when available, but probabilistic reproducibility may be sufficient when the induced behavior is consistently observed under defined conditions.

\subsection{Step 6: Report Penetration Evidence}

The final step is to report the evidence in a form that connects adversarial action to operational consequence. A report that merely states that the model produced an undesirable output is insufficient. The report must show how the adversary influenced the system, what behavior was induced, why the behavior violated an operational objective, and what mitigation is required.

A minimum evidence record for AI-enabled penetration should include the threat model, attacker capability, influence surface, test scenario, induced AI-governed behavior, operational objective violated, success frequency or reproducibility conditions, operational impact, affected controls, and remediation guidance. When human operators are part of the workflow, the report should also document how the AI output shaped human judgment or action. When tools are involved, the report should include tool permissions, tool-call traces, authorization checks, and downstream effects.

\begin{table*}[!t]
\centering
\caption{Minimum evidence elements for objective-driven AI penetration testing.}
\label{tab:ai-pentest-evidence}
\renewcommand{\arraystretch}{1.15}
\begin{tabular}{p{0.22\textwidth} p{0.70\textwidth}}
\hline
\textbf{Evidence element} & \textbf{Purpose} \\
\hline
Threat model & Specifies adversary role, access, capability, constraints, and assumptions. \\
\hline
Operational objective & Defines the mission-level outcome or constraint that must be preserved. \\
\hline
Influence surface & Identifies the resource, interface, input, context, data, tool, memory, or environment used by the adversary. \\
\hline
Adversarial scenario & Describes the sequence of actions used to influence AI-governed behavior. \\
\hline
Induced behavior & Records the prediction, recommendation, generated output, tool call, plan, classification, ranking, or action produced by the system. \\
\hline
Objective violation & Explains how the induced behavior caused, enabled, or materially contributed to failure of the operational objective. \\
\hline
Reproducibility evidence & Provides trial count, success rate, test conditions, model version, configuration, prompts, inputs, retrieval state, or environmental conditions. \\
\hline
Operational impact & Assesses severity in relation to confidentiality, integrity, availability, safety, compliance, mission continuity, or human decision quality. \\
\hline
Remediation guidance & Connects the finding to controls across prompts, models, data, tools, permissions, workflow design, monitoring, and human oversight. \\
\hline
\end{tabular}
\end{table*}
\FloatBarrier
This evidence structure extends conventional penetration-test reporting rather than replacing it. Traditional reports identify vulnerabilities, exploit paths, affected assets, impact, and remediation. Objective-driven AI reports add behavioral susceptibility, influence path, objective violation, and probabilistic evidence. The resulting report allows defenders to distinguish between a general AI quality issue and an adversarially exploitable behavior that threatens the system's operational purpose.

\subsection{Assessment Outcomes}

Objective-driven AI penetration testing can produce several distinct outcomes. The strongest outcome is confirmed AI-enabled penetration: the tester demonstrates a feasible influence path that repeatedly or sufficiently induces behavior violating an operational objective. A second outcome is conditional penetration: the behavior violates an objective only under specific configurations, permissions, retrieval states, or environmental assumptions. A third outcome is behavioral susceptibility: the system shows adversarially influenced behavior, but the observed behavior does not yet violate a defined operational objective. A fourth outcome is non-security model failure: the system behaves incorrectly, but no feasible adversarial path is established. A fifth outcome is no observed penetration under the tested threat model.

These distinctions matter for remediation and prioritization. Confirmed penetration requires risk treatment because adversarial capability, influence path, behavior, and objective violation are all demonstrated. Conditional penetration may require configuration changes, permission reduction, retrieval controls, workflow constraints, or compensating human review. Behavioral susceptibility may require monitoring, additional testing, or narrowing of model authority. Non-security model failure may belong to model-quality assurance rather than penetration testing. No observed penetration does not establish security in general; it establishes only that the tested scenarios did not demonstrate penetration under the stated assumptions.

\subsection{Control Points for Remediation}

Objective-driven testing also changes how mitigations are organized. Controls may be applied at the resource layer, the influence layer, the behavioral layer, or the objective layer. Resource-layer controls include conventional measures such as patching, access control, segmentation, credential protection, secure deployment, supply-chain controls, and monitoring. Influence-layer controls include input validation, content provenance, retrieval filtering, data-quality controls, sensor hardening, and separation of trusted instructions from untrusted data. Behavioral-layer controls include constrained decoding, policy enforcement, tool-use guards, action validation, anomaly detection, sandboxing, and human confirmation for high-impact actions. Objective-layer controls include escalation rules, fail-safe defaults, independent verification, separation of duties, and mission-level monitoring.

The same penetration finding may require controls across multiple layers. An indirect prompt injection that causes an agent to close a security ticket may require retrieval filtering, prompt isolation, tool-permission reduction, confirmation gates, logging, and escalation policy changes. A sensor manipulation that causes unsafe perception behavior may require sensor fusion, environmental robustness testing, anomaly detection, and operational fallback procedures. A poisoning attack against a decision-support system may require data lineage, training-set validation, model monitoring, and human-review thresholds. Objective-driven testing therefore produces remediation guidance that reflects the full path from adversarial influence to operational failure.

\subsection{From Vulnerability Discovery to Behavioral Assurance}

The purpose of objective-driven AI penetration testing is not to replace vulnerability discovery. It is to connect vulnerability discovery to behavioral assurance. A system may have well-protected infrastructure yet remain susceptible to adversarial behavioral induction. Conversely, a model may produce occasional errors without creating a feasible penetration path. The relevant security question is whether an adversary can exploit the relationship among exposed interfaces, AI-governed behavior, and operational objectives.

This framing makes penetration testing more aligned with how AI-enabled systems are used in practice. AI components increasingly summarize, rank, recommend, classify, plan, and act within operational workflows. Their security cannot be evaluated solely by inspecting the resources that host them or the accuracy of the models in isolation. It must be evaluated by testing whether adversaries can steer the system's behavior away from the outcomes it is supposed to preserve. Objective-driven AI penetration testing provides a structured way to make that evaluation explicit, evidence-based, and actionable.

\section{Running Example: SOC Assistant Manipulation}
\label{sec:soc-assistant-example}

Consider an AI-enabled security operations center (SOC) assistant deployed to support incident triage. The assistant receives alerts from a security information and event management platform, summarizes logs and endpoint telemetry, retrieves threat-intelligence reports, recommends severity levels, drafts analyst notes, and invokes predefined response playbooks when escalation criteria are met. The system is not fully autonomous: destructive actions require human approval, but the assistant can influence analyst attention by ranking incidents, suppressing duplicates, recommending closure, and preparing escalation summaries.

The operational objective of the system is to preserve accurate and timely incident triage. More specifically, high-severity incidents must not be downgraded, closed, or omitted from escalation without sufficient evidence and human analyst confirmation. This objective gives security meaning to the assistant's behavior. A wrong summary is not automatically a penetration outcome. A wrong summary becomes security-relevant when it causes or materially contributes to failure of incident escalation under an adversarially feasible path.

\subsection{System and Trust Boundaries}

The assistant operates over several inputs and resources. Trusted inputs include internal detection rules, authenticated analyst commands, approved playbook definitions, and organization-approved policy documents. Untrusted or partially trusted inputs include raw logs, endpoint events, user-submitted ticket comments, external threat-intelligence webpages, email headers, URLs, sandbox reports, and retrieved open-source intelligence. The assistant also has access to tools, including ticket update functions, case-prioritization functions, search over historical incidents, enrichment APIs, and playbook recommendation modules.

The central security boundary is not only between authenticated and unauthenticated access. It is also between trusted instructions and untrusted content. The assistant may be instructed by its system prompt to prioritize verified alerts, cite retrieved evidence, and escalate high-severity incidents. At the same time, it may process untrusted text embedded in logs, webpages, emails, or ticket comments. This creates an instruction--data ambiguity: content that should be treated as data may be interpreted by the model as an instruction. In an agentic workflow, this ambiguity can interact with tool access and produce a confused-deputy pattern, where the assistant uses legitimate authority in a way that serves the adversary's objective \cite{hardyConfusedDeputy,owaspLLMTop10}.

\subsection{Threat Model}

The adversary is an external attacker with no credentials to the SOC platform, no access to the assistant's system prompt, no ability to modify model weights, and no access to internal playbook definitions. The adversary can, however, influence artifacts that the SOC assistant may later process. These artifacts may include a malicious domain, webpage, log string, email header, phishing payload, incident artifact, or ticket comment generated as part of an intrusion attempt. This capability is realistic for many SOC workflows because detection systems routinely ingest attacker-controlled strings, URLs, filenames, command-line arguments, HTTP headers, and external enrichment content.

The adversary's objective is not to compromise the assistant's infrastructure. The objective is to reduce the likelihood that a high-severity incident is escalated. The adversary attempts to induce the assistant to misclassify the incident as benign, omit relevant indicators of compromise, recommend closure, or delay escalation to a human analyst. The attack is successful only if the induced behavior violates the operational objective of accurate and timely incident triage.

\subsection{Adversarial Influence Path}

The influence path begins when the adversary places an instruction-like payload in content that the assistant later observes. For example, the adversary may register a domain containing a webpage that appears to be a benign software-update page but includes hidden or visible text instructing automated systems to treat related alerts as false positives. Alternatively, the adversary may embed instruction-like text in a command-line argument, filename, email header, or ticket comment. The malicious text is not code execution. It is not credential theft. It is not direct modification of the assistant. It is adversarial content placed in the assistant's information environment.

During triage, the SOC assistant retrieves or summarizes the affected artifact. The assistant's prompt requires it to analyze the incident, assess severity, and recommend whether escalation is required. The adversarial content attempts to override this task by instructing the assistant to downgrade the alert, exclude certain indicators, or avoid escalation. If the assistant treats the untrusted content as an instruction rather than evidence, the adversary influences AI-governed behavior through an ordinary data path.

The resulting chain can be summarized as follows:

\begin{quote}
Adversarial content $\rightarrow$ retrieved or ingested evidence $\rightarrow$ assistant reasoning and summarization $\rightarrow$ severity recommendation or ticket action $\rightarrow$ failure of incident triage.
\end{quote}

This is an AI-enabled penetration path because the decisive failure occurs in the behavioral layer. The adversary has not necessarily compromised the SOC platform, the LLM service, the model weights, the ticketing system, or the analyst account. The adversary has instead induced the assistant to behave in a way that violates the operational objective.

\subsection{Behavioral Failure Criteria}

The test defines behavioral failure before execution. The assistant is considered to have violated the operational objective if one or more of the following behaviors occur under the adversarial scenario: a high-severity incident is downgraded below the escalation threshold; relevant indicators of compromise are omitted from the analyst-facing summary; the assistant recommends closure despite evidence of compromise; the assistant invokes or recommends a non-escalation workflow when escalation criteria are met; or the assistant presents adversarial content as authoritative evidence for reducing severity.

These criteria distinguish penetration from ordinary model imperfection. A minor wording error in the summary does not count as penetration. A harmless hallucination unrelated to escalation does not count as penetration. A severity recommendation that differs from an analyst's preference does not count as penetration unless it violates the defined escalation policy. The failure must be both adversarially induced and operationally meaningful.

\subsection{Test Execution}

The test is executed as a controlled scenario. The system is configured with a known alert, a known escalation policy, a known retrieval corpus, and a known assistant configuration. A baseline run is first performed without adversarial content to confirm that the assistant correctly identifies the incident as high severity and recommends escalation. The adversarial content is then introduced through a feasible influence surface, such as a webpage, ticket comment, log field, or retrieved enrichment artifact. The same or equivalent alert is processed again under the adversarial condition.

Because LLM and agentic behavior may be stochastic, the test is repeated across multiple trials. Each trial records the model version, system prompt version, retrieval state, tool permissions, temperature or decoding configuration, alert content, retrieved artifacts, assistant output, tool calls, severity recommendation, and final ticket state. The relevant evidence is not only whether one output is harmful, but whether the adversary can induce objective-violating behavior with sufficient consistency under the defined conditions.

\subsection{Penetration Evidence}

A confirmed penetration finding requires a complete evidentiary chain. The report must show the adversary capability, the influence surface, the adversarial content, the induced assistant behavior, the operational objective violated, and the reproducibility conditions. Table~\ref{tab:soc-example-evidence} illustrates how the evidence can be structured.

\begin{table*}[!t]
\centering
\caption{Evidence structure for the SOC assistant manipulation example.}
\label{tab:soc-example-evidence}
\renewcommand{\arraystretch}{1.15}
\begin{tabular}{p{0.22\textwidth} p{0.70\textwidth}}
\hline
\textbf{Element} & \textbf{SOC assistant example} \\
\hline
Operational objective & High-severity incidents must not be downgraded, closed, or omitted from escalation without sufficient evidence and human analyst confirmation. \\
\hline
Threat model & External adversary cannot compromise SOC infrastructure or model weights, but can influence content later ingested or retrieved by the assistant. \\
\hline
Influence surface & Log field, ticket comment, webpage, email header, URL, filename, enrichment report, or other artifact processed during triage. \\
\hline
Adversarial action & Place instruction-like content that attempts to override triage behavior, suppress escalation, or reclassify malicious activity as benign. \\
\hline
AI-governed behavior & Assistant summary, severity recommendation, evidence selection, ticket update, playbook recommendation, or escalation decision. \\
\hline
Objective violation & Assistant downgrades a high-severity incident, omits indicators of compromise, recommends closure, or fails to escalate when escalation criteria are met. \\
\hline
Evidence & Baseline comparison, trial count, success rate, retrieved context, assistant transcript, tool-call logs, ticket state, and analyst-facing output. \\
\hline
Impact & Delayed response, missed escalation, analyst misprioritization, increased dwell time, or failure to contain an active intrusion. \\
\hline
\end{tabular}
\end{table*}
\FloatBarrier
The finding should not be reported as `the model was fooled'' or `the assistant hallucinated.'' Those descriptions are too vague. The security finding is that an adversary with feasible external influence over content processed by the SOC assistant induced AI-governed behavior that caused violation of the incident-triage objective. The penetration evidence is the demonstrated path from adversarial content to operational failure.

\subsection{Why This Counts as Penetration}

Under a resource-centric interpretation, the test may appear ambiguous. The attacker did not bypass authentication, obtain administrative access, execute code, exfiltrate data, or persist inside the SOC environment. The assistant's infrastructure may remain uncompromised. Yet the adversary achieved a meaningful security outcome: the system behaved in a way that undermined incident detection and response. For a SOC assistant, this is not a peripheral failure. It directly defeats the operational purpose of the system.

Under the definition proposed in this article, the case is straightforward. The adversary has a feasible influence path through untrusted content. The influence path induces AI-governed behavior in the form of a misleading summary, incorrect severity recommendation, or inappropriate workflow action. That behavior violates the operational objective of accurate and timely incident triage. The result is successful AI-enabled penetration, even in the absence of conventional infrastructure compromise.

This example also shows why AI-enabled penetration testing must preserve conventional discipline. The finding is valid only because the threat model is explicit, the influence surface is feasible, the operational objective is defined, the induced behavior is observable, and the objective violation is evidenced. Without those elements, the scenario would collapse into a generic claim that LLMs can produce bad outputs. With those elements, it becomes a penetration-testing result: an adversary can use an available path to cause mission-relevant security failure.

\subsection{Remediation Implications}

The remediation is not a single model patch. The failure arises from the interaction among retrieval, prompting, tool authority, workflow design, and escalation policy. Effective mitigation may require separating trusted instructions from untrusted evidence, preserving provenance of retrieved content, filtering or labeling external content, constraining tool permissions, requiring human confirmation for severity downgrades, validating summaries against structured alert fields, logging retrieved context and tool calls, and monitoring for instruction-like content in untrusted artifacts.

These controls operate at different layers. Some reduce the adversary's ability to influence the assistant. Others constrain the assistant's authority to act on influenced behavior. Others detect when behavior deviates from the operational objective. The example therefore illustrates the broader principle of objective-driven AI penetration testing: remediation must address the full path from adversarial influence to AI-governed behavior to operational-objective violation.

\section{Implications for Practice}
\label{sec:implications}

Defining AI-enabled penetration as adversarially induced violation of operational objectives changes how penetration tests are scoped, executed, reported, and remediated. The change is not a replacement of established penetration-testing practice. It is an extension of that practice to systems in which learned models mediate decisions, recommendations, tool use, or human judgment. The practical implication is that security teams must evaluate both the compromise of computational resources and the manipulation of AI-governed behavior.

\subsection{Scoping Must Include Operational Objectives}

The first implication is that the scope of an AI penetration test must include operational objectives, not only assets and interfaces. Conventional scopes often list hosts, applications, APIs, cloud accounts, network ranges, identities, and data stores. These remain necessary, but they are insufficient for AI-enabled systems. A complete scope must also identify what the AI component is expected to preserve or prevent in operational terms.

For a SOC assistant, the objective may be accurate incident prioritization and escalation. For a biometric access-control system, it may be rejection of unauthorized users. For a medical decision-support system, it may be avoidance of unsafe recommendations under defined clinical constraints. For an autonomous inspection system, it may be detection and escalation of safety-critical defects. For an AI agent connected to enterprise tools, it may be prevention of unauthorized workflow actions. These objectives determine which AI behaviors matter for security and which failures count as penetration outcomes.

This changes the scoping conversation. The test authorization must specify not only which resources may be tested, but also which operational behaviors may be challenged. It must define whether the tester may inject prompts, modify retrieved documents, manipulate sensor inputs, create synthetic tickets, poison test data, trigger tool calls, or simulate adversarial environmental conditions. Without this behavioral scope, an AI penetration test may remain limited to infrastructure compromise and miss the primary ways in which the AI system can be induced to fail.

\subsection{Threat Models Must Include Influence Capabilities}

The second implication is that threat models must describe adversarial influence capabilities. In conventional testing, threat models often focus on network position, authentication state, privilege level, insider status, physical access, or supply-chain access. AI-enabled systems require these assumptions to be extended to the behavioral layer.

Relevant influence capabilities include the ability to submit prompts, upload documents, create webpages, generate emails, modify ticket comments, affect logs, manipulate sensor inputs, contribute feedback, influence training data, poison retrieval corpora, alter memory, or shape tool outputs. These capabilities may belong to different adversary classes. A public user may influence prompts or uploaded files. A remote attacker may influence logs, URLs, headers, or external webpages. An insider may influence knowledge bases, labels, feedback, or workflow records. A supply-chain adversary may influence model dependencies, plugins, datasets, or third-party tools. A physical adversary may influence sensors, images, audio, lighting, location, or environmental context.

The relevant question is whether the adversary can reach the behavior-producing part of the system through a feasible path. This makes threat modeling more precise. It prevents unrealistic claims based on arbitrary model manipulation, while also preventing underestimation of attacks that operate through ordinary exposed interfaces. An AI penetration test should therefore document the adversary's influence capability with the same care that conventional tests document authentication level, network access, and privilege assumptions.

\subsection{Reports Must Distinguish Vulnerabilities from Behavioral Susceptibilities}

The third implication is that AI penetration-test reports need an expanded finding structure. Conventional findings commonly identify a vulnerability, affected asset, exploit path, evidence, impact, and remediation. AI-enabled findings must also document the behavioral susceptibility that connects adversarial influence to objective violation.

A behavioral susceptibility is a condition under which adversarial input, context, data, tool output, memory, or environmental manipulation can steer AI-governed behavior outside the acceptable envelope for the system's objective. This may involve a prompt-injection weakness, unsafe tool delegation, excessive agency, untrusted retrieval, missing provenance, insufficient confirmation, poor separation of instructions and data, vulnerable sensor processing, weak data-lineage controls, or inadequate monitoring of model behavior.

The reporting distinction matters because not every undesirable output is a security finding. A model-quality issue may require retraining or evaluation, but it is not necessarily penetration. A penetration finding requires a feasible adversarial path, induced AI-governed behavior, and violation of an operational objective. Reports should therefore avoid vague labels such as `the model was fooled'' or `the assistant hallucinated.'' The finding should state the adversary capability, influence surface, induced behavior, affected objective, reproducibility conditions, operational impact, and remediation.

\begin{table}[!t]
\centering
\caption{Extension of conventional penetration-test reporting for AI-enabled systems.}
\label{tab:reporting-extension}
\renewcommand{\arraystretch}{1.15}
\begin{tabular}{p{0.15\textwidth} p{0.39\textwidth} p{0.39\textwidth}}
\hline
\textbf{Report element} & \textbf{Conventional emphasis} & \textbf{AI-enabled extension} \\
\hline
Target & Host, application, API, account, database, or network segment & AI-mediated workflow, model behavior, tool use, retrieval path, sensor channel, or decision loop \\
\hline
Vulnerability & Software defect, misconfiguration, exposed service, weak credential, or access-control flaw & Behavioral susceptibility, prompt/context weakness, unsafe tool delegation, data-flow weakness, retrieval poisoning, or sensor manipulation path \\
\hline
Exploit path & Steps leading to unauthorized access, privilege escalation, disclosure, disruption, or persistence & Influence path leading from adversarial input, data, context, memory, tool output, or environment to AI-governed behavior \\
\hline
Evidence & Shell access, data access, privilege change, service disruption, or resource state change & Transcript, prompt, retrieved context, tool-call trace, model output, behavioral observation, trial count, success rate, and objective violation \\
\hline
Impact & Confidentiality, integrity, availability, business risk, lateral movement, or operational disruption & Operational-objective violation, unsafe decision, missed escalation, unauthorized action, distorted prioritization, or harmful human-AI interaction \\
\hline
Reproducibility & Deterministic exploit steps and affected configurations & Deterministic or probabilistic reproduction under defined model version, prompt, retrieval, tool, memory, and environmental conditions \\
\hline
Remediation & Patch, harden, segment, restrict access, rotate credentials, monitor, or update controls & Constrain AI authority, separate trusted instructions from untrusted data, validate retrieval, restrict tools, add confirmation gates, monitor behavior, and enforce objective-level controls \\
\hline
\end{tabular}
\end{table}
\FloatBarrier
\subsection{Evidence May Be Probabilistic}

The fourth implication is that AI penetration evidence may be probabilistic. Conventional exploit evidence often aims for deterministic reproducibility: given the same vulnerable condition and exploit steps, the same compromise occurs. AI-enabled systems may not behave this way. Outputs may vary with sampling parameters, prompt phrasing, model version, retrieval order, memory state, environmental conditions, or human interaction. This does not remove the need for rigor. It changes the form of evidence required.

Reports should document the conditions under which the behavior was observed. These conditions may include model version, system prompt, application prompt, temperature or decoding configuration, retrieval corpus, retrieved documents, memory contents, tool permissions, input artifacts, sensor conditions, trial count, and success frequency. A finding based on one anomalous output is weak unless the objective violation is severe and the adversarial path is clear. A finding based on repeated objective violation under controlled conditions is stronger, even if the success rate is below one hundred percent.

Severity should account for both probability and consequence. A low-probability behavioral failure may still be unacceptable if the system has high authority or if the operational consequence is severe. Conversely, a frequent behavioral deviation may be lower severity if downstream controls reliably prevent objective violation. AI penetration testing therefore requires evidence that connects behavioral frequency to operational risk rather than treating every failure as equivalent.

\subsection{Controls Must Be Applied Across Layers}

The fifth implication is that remediation must address the full path from adversarial influence to operational failure. In conventional penetration testing, remediation often focuses on the vulnerable asset or control: patch the service, fix the configuration, restrict access, rotate credentials, or improve monitoring. In AI-enabled systems, the weakness may be distributed across prompts, retrieval, memory, model behavior, tool authorization, workflow design, and human oversight.

Controls can be organized across four layers. Resource-layer controls protect the infrastructure that hosts and supports the AI system, including identities, APIs, model repositories, deployment pipelines, data stores, containers, and supply chains. Influence-layer controls reduce the adversary's ability to inject or manipulate the inputs that shape AI behavior, including input validation, provenance tracking, retrieval filtering, content labeling, data-lineage controls, sensor hardening, and separation of trusted instructions from untrusted data. Behavioral-layer controls constrain or monitor AI-governed behavior, including policy enforcement, tool-use guards, action validation, sandboxing, anomaly detection, constrained autonomy, and model-output verification. Objective-layer controls ensure that high-impact decisions remain aligned with mission requirements, including human confirmation, escalation thresholds, separation of duties, fail-safe defaults, independent verification, and operational monitoring.

\begin{table}[!htb]
\centering
\caption{Control layers for mitigating AI-enabled penetration paths.}
\label{tab:control-layers}
\renewcommand{\arraystretch}{1.15}
\begin{tabular}{p{0.15\textwidth} p{0.38\textwidth} p{0.38\textwidth}}
\hline
\textbf{Layer} & \textbf{Primary question} & \textbf{Representative controls} \\
\hline
Resource layer & Are the computational assets protected? & Access control, patching, segmentation, credential protection, secure deployment, supply-chain controls, logging, and infrastructure monitoring. \\
\hline
Influence layer & Can untrusted input, data, context, or environment steer behavior? & Input validation, provenance, retrieval filtering, data-lineage controls, sensor hardening, prompt isolation, trusted/untrusted content separation, and memory controls. \\
\hline
Behavioral layer & Can AI-governed behavior exceed safe or authorized bounds? & Tool-use guards, policy enforcement, constrained autonomy, output validation, anomaly detection, sandboxing, action verification, and behavioral monitoring. \\
\hline
Objective layer & Can induced behavior violate the system's mission? & Human confirmation, escalation rules, independent verification, fail-safe defaults, separation of duties, impact thresholds, and mission-level monitoring. \\
\hline
\end{tabular}
\end{table}

Layered remediation prevents a narrow fix from being mistaken for risk reduction. For example, filtering a malicious prompt pattern may not address the broader problem if the agent still treats untrusted retrieved content as authoritative instruction. Restricting tool permissions may reduce impact but leave the influence surface exposed. Adding human review may reduce objective violation but not explain why the assistant accepted the adversarial content. Effective mitigation addresses the chain, not only the symptom.

\subsection{AI Red Teaming and Penetration Testing Must Be Connected}

The sixth implication is that AI red teaming and penetration testing should not remain separate activities. AI red teaming often evaluates harmful outputs, unsafe behavior, policy bypass, jailbreaks, prompt injection, or misuse scenarios. Penetration testing evaluates adversarial exploitability, impact, and remediation under a defined threat model. AI-enabled systems require these disciplines to converge.

A red-team scenario becomes a penetration-test finding when it establishes a feasible adversarial influence path and connects the induced behavior to operational-objective violation. Conversely, a penetration test of an AI-enabled system is incomplete if it examines only infrastructure and ignores behavioral influence. The combined practice should preserve the rigor of penetration testing while incorporating the scenario diversity and behavioral focus of AI red teaming.

This convergence also changes team composition. Effective AI penetration testing may require expertise in application security, cloud security, identity and access management, adversarial ML, prompt and retrieval design, model evaluation, data governance, human factors, and the operational domain in which the system is deployed. A tester who understands only infrastructure may miss behavioral influence paths. A tester who understands only model behavior may miss conventional compromise paths. The practical unit of assessment is the AI-enabled system as deployed, not the model in isolation.

\subsection{Testing Must Be Integrated into Development and Operations}

The seventh implication is that AI penetration testing should be integrated into development and operational workflows. AI-enabled systems change over time. Models are updated, prompts are revised, retrieval corpora change, tools are added, permissions are adjusted, memory accumulates, user behavior evolves, and environmental conditions shift. A point-in-time test can identify important failures, but it cannot provide continuing assurance if the behavior-producing system changes after the test.

Development pipelines should include adversarial scenario tests for critical operational objectives. Deployment pipelines should validate prompt changes, retrieval changes, tool-permission changes, and model-version changes against known behavioral failure criteria. Operations teams should monitor for objective-relevant behavioral drift, suspicious tool use, unexpected retrieval dependence, repeated policy bypass attempts, and changes in failure frequency. Incident-response processes should treat AI-induced objective violation as a security event when it arises through adversarial influence.

This integration connects AI penetration testing to MLOps, SecOps, and governance. Model evaluation alone is insufficient because deployment context determines the operational consequence of behavior. Infrastructure testing alone is insufficient because behavior may be manipulated without infrastructure compromise. Continuous assurance requires monitoring both the state of resources and the behavior those resources produce.

\subsection{Accountability Requires Clear Boundaries of Authority}

The eighth implication concerns authority. Many AI-enabled failures become severe only when the system has authority to act, prioritize, suppress, recommend, or influence human decisions. A chatbot that produces a bad answer may be low impact in isolation. The same model embedded in a ticketing workflow, access-control system, clinical decision process, financial review pipeline, or incident-response workflow may create a penetration path if adversarial influence changes an operational outcome.

Testing should therefore identify the authority boundary of each AI component. The authority boundary defines what the AI system can do directly, what it can recommend, what it can suppress, what tools it can call, what records it can modify, what humans are likely to accept, and what actions require confirmation. Excessive authority increases the severity of behavioral susceptibility. Clear authority boundaries make objective-driven testing and remediation more precise.

Human oversight should also be treated as a control with failure modes. A human-in-the-loop design does not automatically prevent penetration if the AI system shapes what the human sees, ranks, summarizes, omits, or recommends. Human confirmation is stronger when the human receives independent evidence, sees provenance, understands uncertainty, and has time and authority to challenge the AI output. It is weaker when the AI output becomes the primary basis for judgment.

\subsection{Toward Operational Assurance}

The practical aim of AI penetration testing is operational assurance. The question is not merely whether the model is robust, whether the infrastructure is hardened, or whether a policy exists. The question is whether an adversary can cause the deployed AI-enabled system to violate the purpose for which it is trusted. This requires security teams to evaluate the chain from adversarial influence to AI-governed behavior to operational consequence.

The resulting practice remains grounded in the discipline of conventional penetration testing: explicit threat models, adversarial evidence, impact assessment, reproducibility, and remediation. Its extension is the recognition that, in AI-enabled systems, the decisive security failure may occur at the behavioral layer. A mature assessment must therefore protect resources, test influence paths, observe induced behavior, and measure whether operational objectives remain intact under adversarial conditions.

\section{Open Research Challenges}
\label{sec:open-challenges}

Objective-driven AI penetration testing raises several research challenges that are not fully addressed by existing penetration-testing methodologies, adversarial machine-learning benchmarks, or AI risk-management frameworks. These challenges concern how operational objectives should be specified, how behavioral penetration should be measured, how test scenarios should be generated, how evidence should be standardized, and how testing should be integrated into continuously changing AI-enabled systems.

\subsection{Specifying Operational Objectives}

The first challenge is the formal specification of operational objectives. Conventional penetration testing can often rely on well-established security properties such as confidentiality, integrity, availability, authorization, and isolation. AI-enabled systems require additional objective definitions that are closer to the mission of the deployed system. These objectives may involve correct escalation, safe recommendation, reliable prioritization, policy-compliant tool use, accurate perception, calibrated decision support, or prevention of harmful human-AI interaction.

Specifying such objectives is difficult because they are often contextual, domain-dependent, and partly normative. A safe recommendation in one setting may be unsafe in another. A tolerable false positive rate in one workflow may be unacceptable in another. A model output that is harmless in an advisory interface may become severe when connected to tools or automation. Research is needed on objective-specification languages, policy representations, assurance cases, and domain-specific templates that allow operational objectives to be expressed in testable form.

A useful specification must identify not only the desired outcome, but also the acceptable behavioral envelope around that outcome. It should define which actions are prohibited, which decisions require confirmation, which information must not be omitted, which sources are authoritative, which uncertainty thresholds require escalation, and which deviations constitute security-relevant failure. Without such specifications, AI penetration testing risks confusing general model imperfection with adversarially induced mission failure.

\subsection{Measuring Probabilistic Penetration}

The second challenge is measurement. Conventional penetration testing often seeks reproducible exploit chains. AI-enabled systems may instead produce probabilistic, context-dependent, and state-dependent behavior. An adversarial prompt, poisoned retrieval document, manipulated sensor input, or crafted tool output may induce failure in some trials but not others. The question is how much evidence is sufficient to claim successful penetration.

Research is needed on statistical measures for behavioral penetration. Such measures should account for trial count, success rate, severity of objective violation, model variance, environmental conditions, and operational consequence. A penetration finding against a low-impact assistant may require repeated success across many trials. A finding against a high-authority agent or safety-critical system may be significant even at lower success probability. The evidence standard should therefore combine likelihood and consequence rather than treating penetration as purely binary.

This also requires methods for defining meaningful test populations. A success rate measured over arbitrary prompts or synthetic examples may not reflect real operational exposure. Scenario distributions should represent plausible adversary behavior, realistic inputs, and deployment-specific workflows. Without careful sampling, probabilistic evidence may either exaggerate risk or miss rare but severe objective violations.

\subsection{Generating Realistic Adversarial Scenarios}

The third challenge is adversarial scenario generation. Existing vulnerability scanners and fuzzers are optimized for software defects, protocol behavior, input parsing, and known vulnerability patterns. AI-enabled systems require scenarios that exercise prompts, retrieval pipelines, memory, sensor inputs, tool use, planning, feedback loops, and human-AI interaction. These scenarios must be realistic enough to reflect adversary capability and diverse enough to expose behavioral susceptibility.

Automated scenario generation is especially difficult for agentic AI systems. A successful attack may require multiple steps: placing malicious content, triggering retrieval, influencing reasoning, causing a tool call, bypassing a confirmation rule, and shaping the final human-facing output. Each step may depend on system state, context, permissions, and prior model behavior. Research is needed on test harnesses that can explore these multi-step influence paths while preserving explicit threat-model assumptions.

Scenario generation also requires domain knowledge. A generic prompt-injection test may not reveal whether a SOC assistant violates incident-triage objectives, whether a clinical assistant violates care constraints, or whether an industrial inspection system misses safety-critical defects. Future tools must therefore combine general adversarial techniques with domain-specific objective models, workflow simulators, and operational data.

\subsection{Benchmarking System-Level AI Penetration}

The fourth challenge is benchmarking. Existing adversarial ML benchmarks often evaluate isolated model robustness against defined perturbations or attacks. Such benchmarks are useful but insufficient for objective-driven AI penetration testing. The relevant unit of evaluation is the deployed AI-enabled system: model, prompts, retrieval, tools, data flows, user interface, permissions, monitoring, and human workflow.

A benchmark for AI penetration testing should include operational objectives, threat models, adversarial influence surfaces, realistic workflows, observable behavior, and objective-violation criteria. It should evaluate not only whether a model output changes, but whether the system's behavior crosses a mission-relevant threshold. For agentic systems, benchmarks should include tool permissions, memory, retrieval state, and multi-step task execution. For cyber-physical systems, they should include environmental variation, sensor conditions, and downstream control or decision consequences.

Benchmark design must also avoid rewarding unrealistic attacks. A benchmark that assumes arbitrary prompt access, unconstrained tool permissions, or direct modification of model state may overstate risk for some deployments. Conversely, a benchmark that tests only isolated prompts may understate risk in systems where adversaries can influence retrieved content, logs, emails, images, or external webpages. The benchmark must therefore encode adversary capability and deployment assumptions explicitly.

\subsection{Standardizing Evidence and Reporting}

The fifth challenge is standardization of evidence. Conventional penetration-test reports have established patterns for documenting vulnerabilities, affected assets, exploit steps, impact, and remediation. AI-enabled penetration requires additional evidence fields: operational objective, adversarial influence surface, induced behavior, objective violation, test conditions, stochastic variability, success frequency, model configuration, retrieval state, memory state, tool permissions, and human-interaction effects.

Without reporting standards, AI penetration findings will be difficult to compare, reproduce, or prioritize. One report may treat a single harmful output as a critical vulnerability, while another may dismiss repeated behavioral manipulation as a model-quality issue. Research is needed on reporting schemas, severity models, and evidence thresholds that distinguish model error, policy failure, behavioral susceptibility, and confirmed AI-enabled penetration.

Severity scoring is particularly unresolved. Existing vulnerability scoring systems are not designed to capture the full path from adversarial influence to AI-mediated objective violation. A useful severity model should consider adversary capability, exposure of the influence surface, reliability of induction, authority of the AI component, reversibility of the action, human oversight, operational consequence, and availability of compensating controls. It should also distinguish between direct unauthorized action and indirect manipulation of human decision-making.

\subsection{Integrating Testing with MLOps and SecOps}

The sixth challenge is continuous integration of AI penetration testing into development and operations. AI-enabled systems change frequently. Model versions are updated, prompts are revised, retrieval corpora evolve, tools are added, permissions change, memory accumulates, and users adapt their behavior. A penetration test performed at one point in time may become stale after a model update, retrieval change, or workflow modification.

Research is needed on integrating objective-driven adversarial tests into MLOps and SecOps pipelines. Prompt changes, model updates, retrieval-index changes, tool-permission changes, and policy updates should be tested against known behavioral failure criteria before deployment. Runtime monitoring should detect objective-relevant behavioral drift, suspicious tool use, abnormal retrieval dependence, repeated instruction-conflict events, and changes in success rates for adversarial scenarios.

This integration also raises operational questions. Continuous AI penetration testing may require synthetic adversarial data, safe test environments, red-team automation, shadow deployments, and feedback loops from production incidents. It must avoid interfering with live operations while still reflecting realistic system behavior. The research challenge is to make AI penetration testing repeatable, safe, and operationally meaningful across the lifecycle of the system.

\subsection{Human-AI Interaction as an Attack Surface}

The seventh challenge is the role of human operators. Many AI-enabled systems do not act autonomously, but they shape human judgment through summaries, rankings, recommendations, alerts, explanations, or confidence scores. In these systems, objective violation may occur because adversarially influenced AI output changes what a human sees, trusts, prioritizes, or ignores.

This creates a security problem that is not fully captured by model evaluation or infrastructure testing. A technically incorrect output may be harmless if the human operator detects and corrects it. A plausible but misleading output may be severe if it causes the operator to miss a critical event. Research is needed on how to test human-AI decision loops under adversarial influence, how to measure manipulation of attention or trust, and how to design oversight mechanisms that remain effective when the AI system itself controls evidence presentation.

Human-in-the-loop designs should therefore be evaluated as security controls, not assumed to be sufficient. The strength of the control depends on provenance, independent evidence, explanation quality, workload, time pressure, interface design, and the operator's ability to challenge the AI recommendation. Objective-driven AI penetration testing requires methods for evaluating these factors without reducing human judgment to a passive final checkpoint.

\subsection{Toward a Science of Behavioral Penetration}

The broader challenge is to develop a science of behavioral penetration for AI-enabled systems. Such a science must connect adversary models, AI behavior, system architecture, operational objectives, human interaction, and evidence standards. It must preserve the rigor of conventional penetration testing while accommodating stochastic behavior, contextual influence, and mission-level consequences.

Progress will require collaboration across cybersecurity, machine learning, human-computer interaction, safety engineering, systems engineering, and domain-specific operational communities. The goal is not merely to catalogue more AI attacks. The goal is to determine when adversarial influence over AI-governed behavior constitutes penetration, how such penetration can be tested, how evidence should be interpreted, and how systems can be designed to withstand it.

\FloatBarrier
\section{Conclusion}
\label{sec:conclusion}

AI-enabled systems change the meaning of penetration because they change how security-relevant behavior is produced. In conventional systems, adversarial success is often evidenced through unauthorized access, privilege escalation, data exposure, persistence, disruption, or control over computational resources. These forms of compromise remain essential to security testing, and they remain fully relevant to AI-enabled systems. However, they no longer exhaust the ways in which an adversary can produce meaningful security failure.

When learned models mediate decisions, recommendations, classifications, tool calls, plans, or human judgment, adversarial influence may operate through behavior rather than direct infrastructure compromise. A prompt, retrieved document, sensor input, poisoned example, memory entry, tool output, or manipulated context may steer the system toward an outcome that violates its operational purpose. In such cases, the decisive question is not only whether the attacker got into the system. It is whether the attacker made the system act against the objective it was deployed to preserve.

This article has argued that penetration testing for AI-enabled systems should therefore be extended from resource-centric compromise to objective-driven behavioral evaluation. The proposed definition treats AI-enabled penetration as the feasible induction of AI-governed behavior that violates one or more operational objectives under an explicit threat model. This definition preserves the value of conventional penetration testing while making room for adversarial pathways that operate through AI-mediated behavior. Resource compromise remains one path to penetration, but it is not the only one.

The practical consequence is that AI penetration testing must begin with operational objectives, map the AI-governed behaviors that support them, identify adversarial influence surfaces, define behavioral failure criteria, execute scenario-based tests, and report evidence that links adversarial action to objective violation. Reports must distinguish confirmed penetration from general model failure, behavioral susceptibility, and non-security quality defects. They must also account for probabilistic evidence, tool authority, retrieval state, model configuration, human oversight, and downstream operational impact.

The broader lesson is that security assessment must follow the locus of control. As AI components acquire greater influence over workflows, decisions, and actions, the security boundary expands from protected resources to the behavior those resources enable. A system whose infrastructure remains intact may still be penetrated if an adversary can reliably induce AI-governed behavior that defeats its mission. Future penetration testing must therefore ask two questions together: can the attacker compromise the system's resources, and can the attacker make the system violate its operational objectives?

\section*{Acknowledgements}

The authors acknowledge the use of ChatGPT, developed by OpenAI, as a writing-support tool during the preparation of this manuscript. Its use was limited to language polishing, improving readability, and refining the presentation of the text. The authors reviewed, edited, and verified the manuscript, including its technical arguments, definitions, references, and conclusions. The authors take full responsibility for the content, accuracy, integrity, and final form of the paper.

\bibliography{references}

\end{document}